%% file: letter_3.tex
\documentclass[reprint,prl,twocolumn,superscriptaddress,floatfix]{revtex4}

\usepackage{bm}
\usepackage{graphicx}%
\usepackage{epsf}
\newcommand{\ba}{\begin{eqnarray*}}
\newcommand{\ea}{\end{eqnarray*}}
\newcommand{\be}{\begin{equation}}
\newcommand{\ee}{\end{equation}}
\newcommand{\bd}{\begin{displaymath}}
\newcommand{\ed}{\end{displaymath}}

\newcommand{\plotsize}{0.40\textwidth}
\newcommand{\plotgap}{0.04\textwidth}
\newcommand{\plotangle}{0}

\input{./adjustment.tex}

\begin{document}

\title{Determination of $\Delta$ resonance parameters from lattice QCD}

\date{\today}

\author{C. Alexandrou}

\affiliation{Department of Physics, University of Cyprus, P.O. Box 20537, 1678 Nicosia, Cyprus}
\affiliation{Computation-based Science and Technology Research Center, Cyprus Institute, 20 Kavafi Str., 2121 Nicosia, Cyprus}

\author{J.~W.~Negele}

\affiliation{Center for Theoretical Physics, Laboratory for Nuclear Science and 
        Department of Physics, Massachusetts Institute of Technology, Cambridge, Massachusetts 02139, U.S.A.}

\author{M. Petschlies}

\affiliation{Computation-based Science and Technology Research Center, Cyprus Institute, 20 Kavafi Str., 2121 Nicosia, Cyprus}

\author{A. Strelchenko}

\affiliation{Scientific Computing Division, Fermilab, P.O. Box 500, Batavia, IL 60510-5011}

\author{A. Tsapalis}

\affiliation{Hellenic Naval Academy, Hatzikyriakou Ave., Pireaus 18539,
  Greece and Department of Physics, National Technical University of
  Athens, Zografou Campus 15780, Athens, Greece}

\begin{abstract}
A method suitable for extracting resonance  parameters of unstable
 baryons in lattice QCD is examined.  The method is applied 
to the strong decay of the $\Delta$ 
to a pion-nucleon state, extracting the $\pi N\Delta$ coupling constant and $\Delta$ decay width.

\end{abstract}

\maketitle

\section{Introduction}
The investigation of resonances and hadronic decays
using the underlying theory of the strong interactions, Quantum
Chromodynamics (QCD), is of fundamental importance for nuclear and particle physics. 
 Recent progress in the simulation of QCD on the lattice opens up the possibility of understanding from first principles the phenomenology of the hadronic spectra such
as the Roper resonance, the pattern and nature of the other observed resonances,  and identification of 
exotic states. Unlike the study of stable low lying hadrons where
the lattice QCD formalism is well developed and most suitable, the investigation
of resonances and decays is intrinsically more difficult and it is only very recently that numerical results mainly on the width of mesons have emerged.  
  The basic difficulty lies in the
fact that scattering states  cannot be realized on a lattice 
with finite volume in Euclidean field theory. 
One approach to calculate the decay width of resonances
 is to make use of the dependence of the energy of interacting particles in 
the finite volume of the lattice. The energy shift
in finite volume  can be  related to
the scattering parameters and in particular to the decay width~\cite{Luscher:1985dn,Luscher:1986pf}.
 This approach has been successfully applied 
mostly to calculate the width of the $\rho$-resonance and
the scattering length of $\pi-\pi$ and other two-meson systems~\cite{Torok:2009dg,Feng:2010es,Li:2007ey}.

An alternative approach was proposed in Ref.~\cite{McNeile:2002az}. This approach
 is based on  the  mixing 
of hadronic states on the lattice when their energies are close and the evaluation of 
the corresponding
transition matrix element. 
This approach has been shown to work for the decay of the lightest vector meson on the lattice, $\rho \to \pi\,\pi$
~\cite{McNeile:2002fh} as well as for the $B$-meson~\cite{McNeile:2004rf}.
In this letter we  apply, for the first time, this method to hadronic decays in the baryon sector and show its applicability in the well-known case of
the $\Delta$ decay to a pion-nucleon state.

Specifically, we choose the isospin channel $\Deltapp \to \pi^+\,p$.
Thus we seek an evaluation of the $\Delta$ to $\pi N$ transition amplitude
$\langle{\Delta,\,\vec{Q},\, \vec{q},\,t_f|\,\pi\,N,\,\vec{Q},\,\vec{q},\,t_i}\rangle$ where $t_f\,(t_i)$ is the final (initial) time.

\begin{figure}[h]
\includegraphics[width=\plotsize,angle=\plotangle]{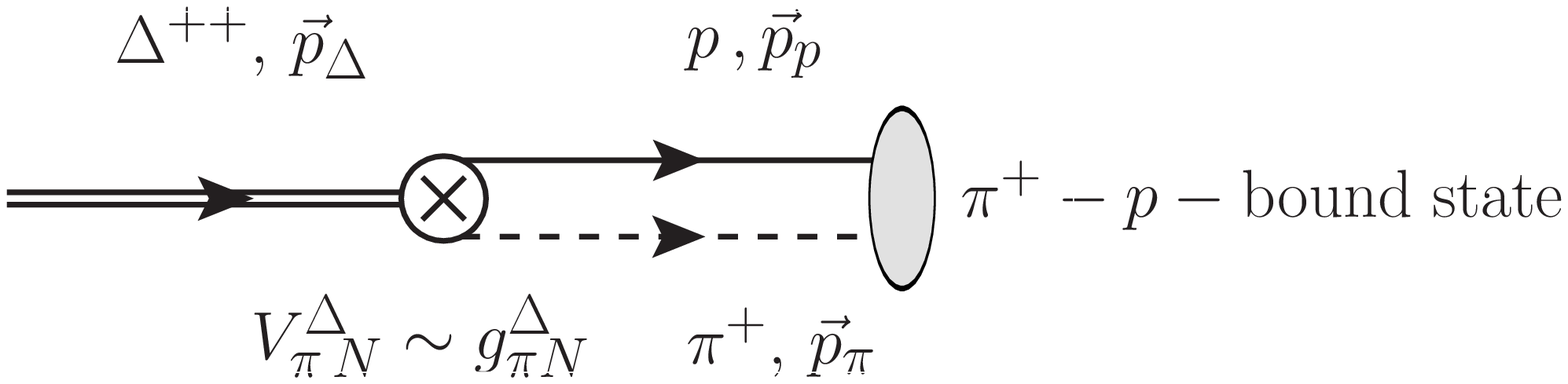}\hspace{\plotgap}
\caption{}
\label{}
\end{figure}

We consider a two-state transfer matrix $\mathrm{T}$, which parameterizes the 
transition amplitude $x= \langle \Delta  |\pi N\rangle $ of the form
\ba
\mathrm{T} &= \epow{-a\bar{E}}\,
  \begin{pmatrix}
    \epow{-a\delta/2} & ax \\ ax & \epow{+a\delta/2}
  \end{pmatrix}\,,
\ea
where $\bar{E} = ( E_\Delta + E_{\pi N} ) / 2$ and $\delta = E_{\pi N} - E_\Delta$.
Restricting to the  $\Delta$ and $\pi N$ states, the matrix element can be written as
\ba
& &\brackets{\Delta,\,t_f\,|\,\pi N,\,t_i} =\brackets{\Delta\,|\,\epow{-H(t_f-t_i)}\,|\,\pi N}
= \brackets{\Delta\,|\,\mathrm{T}^{n_{fi}}\,|\,\pi N} \\
&=& \sum\limits_{n=0}^{n_{fi}-1}\,\epow{-(\bar{E}-\delta/2)t_n}\,
\brackets{\Delta\,|\,T\,|\,\pi N}\,
\epow{-(\bar{E}+\delta/2)(\Delta_{fi}-t_n-a)}\\
&=& ax\,\frac{\sinh(\delta\,\Delta t_{fi}/2)}{\sinh(a\delta/2)}\,\epow{-\bar{E}\Delta t_{fi}}
\ea
where $a$ is the lattice spacing. 
We construct a ratio of 3-point to
2-point functions that for $t_f - t_i = \Delta t_{fi} \rightarrow \gg 1$ yields the $\Delta$ to   $\pi N$ overlap
\be
R(\Delta t_{fi},\Qvec,\qvec) = \frac{ C^{\Delta \to \pi N}_\mu(\Delta t_{fi},\,\Qvec,\qvec) }{
\sqrt{ C^{\Delta}_\mu(\Delta t_{fi},\,\Qvec) \,C^{\pi N}(\Delta t_{fi},\,\Qvec,\,\qvec)}
}\,,
\label{ratio}
\ee
where 
$C^{\Delta}_{\mu}(\Delta t_{fi},\,\Qvec)$,
$C^{\pi N}(\Delta t_{fi},\,\Qvec,\qvec)$ are the $\Delta$ and $\pi N$ two-point
correlation functions and  
$C^{\Delta \to \pi N}_{\mu}(\Delta t_{fi},\,\Qvec,\qvec)$
is the $\Delta$ to $\pi N$ three-point function.
For the $\Delta$ we use the standard interpolating field
\be
J_{\Delta \mu}^\alpha(t,\xvec) =\epsilon_{abc}\,u^{Ta}(t,\xvec)\, C\gamma_\mu\,u^{b}(t,\xvec) \,u^{\alpha c}(t,\xvec)
\ee
whereas we approximate the $\pi N$ state with the product of the  nucleon and pion interpolating fields given by
\be
J_{\pi N}^\alpha(t,\qvec,\xvec) = \sum\limits_{\yvec}\,J_\pi(t,\yvec+\xvec)\,J_N^\alpha(t,\xvec)\,\epow{-i\qvec\yvec}
\ee
where
$J_\pi(t,\yvec) = \dbar(t,\yvec)\, \gammafive\, u(t,\yvec)$ and
$J_N^\alpha(t,\xvec) = \epsilon_{abc}\,u^{Ta}(t,\xvec)\, C\gammafive\,d^{b}(t,\xvec) \,u^{\alpha c}(t,\xvec)$ are the standard pion and proton interpolating fields.  Moreover, on the level of Wick contractions 
we keep only the dominant contribution, which corresponds to the product
of the individually contracted $\pi- \pi-$current and $N-N-$ current.
For large Euclidean time we expect the interpolating fields to dominantly generate  the $\Delta$ and the
$\pi-N$ state which have overlaps with the vacuum that cancel in the ratio of Eq.~(\ref{ratio}).
 The $\pi N$ state
is constructed to have a relative momentum of $\qvec \ne 0$. In this way the lattice ground state will have overlap
with the two-particle state having orbital angular momentum $l=1$. Together with the nucleon spin $s_N = 1/2$, we
thus expect a dominant coupling $l+s_N \rightarrow J = 3/2 = J_\Delta$, which allows mixing with the  $\Delta$
state.

\section{Lattice QCD Calculation}

We perform a  lattice calculation using a hybrid action
of domain wall (DW) valence quarks and gauge configurations
generated with two degenerate staggered up and down quarks and a  strange 
staggered quark fixed to its physical value ($N_f=2+1$) using the Asqtad improved action~\cite{Bernard:2001av}.
The light quark mass in the simulations corresponds to a lightest pion mass of $\mps \approx 360\mev$. The
spatial length of the lattice is $L = 3.4\fermi$ and the lattice spacing is $a \approx 0.124\fermi$.
The light  quark mass in the hybrid theory is determined
by matching the pion mass to that of the lightest pion generated by the Asqtad action
as described in Ref.~\cite{WalkerLoud:2008bp}. 

For kinematics, we choose $\qvec = \pm\, k\,\hat{e}_i$ with  $k = 2\pi/L$
and $\hat{e}_i$ the unit vector in spatial direction $i=1,\,2,\,3$ and we work in the rest frame of the  $\Delta$  setting the total momentum $\Qvec = 0$. Each choice of a pair $\pm,\,i$ defines a corresponding ratio $R_{\pm, i}$,
that we label with these same indices. We evaluate $R_{\pm, i}$ by analyzing 210 gauge configurations with 4 randomly chosen
source locations per configuration subject to the constraint of maximal distance $T/4$ between neighboring source time-slices.
We optimize ground state dominance by using Gaussian smearing on the interpolating fields and by
performing APE smearing on the gauge field configurations that enter the
Gaussian smearing function.

In Table~\ref{tab:spectrum} we list the energies and the momenta corresponding to our kinematics.
The energy of the pion-nucleon system labeled
``$\pi+N$'' is given by the sum of individual pion and nucleon energies.
This corresponds to  approximating  the two-particle state
as a product of one-particle states.
\begin{center}
\begin{table}
\caption{The energy and momentum of the states considered.}
\begin{tabular}{c|cccc}
state         & $\pi$          & $N$             & $\pi + N$      & $\Delta$      \\
\hline
$|\qvec|$     & $2\pi/L$       & $2\pi/L$        & $2\pi/L$       & 0             \\
\hline
$aE(\qvec)$   & $0.3170\,(09)$ & $0.7547\,(72)$  & $1.0717\,(74)$ & $0.965\,(16)$
\label{tab:spectrum}
\end{tabular}
\end{table}
\end{center}

In order to increase statistics we average over forward and backward propagating pion-nucleon and $\Delta$ as well
as over all momentum directions. From the
values of the mass and energy given in Table~\ref{tab:spectrum}  we estimate
 the energy splitting between the
$\Delta$ and pion-nucleon bound state to be,
\ba
a\delta \approx aE_{\pi+N} - aE_\Delta &=& 0.106 \,(16)\,.
\label{eq:delta_estimate}
\ea
Note that the pion is so heavy that $\delta > 0$ and the $\Delta$
cannot decay since its mass is 
below that of the pion-nucleon bound state. However, for a $\Delta$ energy 
close to the energy of the pion-nucleon system one can evaluate the
overlap $x$ between the $\Delta$ and the pion-nucleon decay channel. 
 Only in the limit  $\Delta t_{fi}\to \infty$ and $a\to 0$ do we recover
the Dirac-$\delta$ function of the continuum theory,
\ba
\delta_{\Delta t_{fi},a}(\delta) &=& a\,\frac{\sinh(\delta\,\Delta t_{fi}/2)}{\sinh(a\delta/2)}
\xrightarrow[a\to 0]{\Delta t_{fi} \to \infty} 2\pi\,\delta(p^0_{\pi N} - p^0_\Delta) \,,
\ea
enforcing exact equality of the $\Delta$ and the pion-nucleon energies.
Based on these observations we use two ans\"atze to fit the ratio $R$ and extract the transition amplitude $B$,
\ba
  f_1(t) &=& A + B\,a\,\frac{\sinh(\delta\,t/2)}{\sin(a\delta/2)}\\
  f_2(t) &=& A + B\,t \: ( + C\,t^3 )\,.
\ea
The first version, $f_1$, corresponds to the expected functional dependence on the lattice given a non-zero
splitting of the energy levels. The form $f_2$ is the linearized version of
$f_1$ and represents
the limit of $f_1$ for $\delta \to 0$,  whereas by adding the term cubic in $t$ we check for the significance of a
potential curvature.  Given the sizable energy gap we observe from the spectral analysis it will
be interesting to check the impact of the splitting and thus the possible curvature on the fit. 

\begin{figure}
\includegraphics[width=\plotsize,angle=\plotangle]{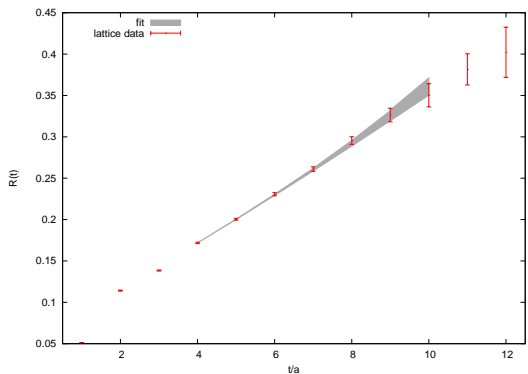}\hspace{\plotgap}
\caption{Ratio for combined forward and backward propagation and averaged over six momentum directions.
  The shaded band shows the fit using $f_1$ in the interval $4 \le t/a \le 10$.}
\label{fig:cmratio_fwb_ravg_uwerr_w_fit_comparison}
\end{figure}

As was done for the two-point functions, we improve the signal of the ratio 
by combining  data from forward and backward propagation and average the
ratio obtained for all six combinations $(\pm ,\,i)$. The resulting ratio
 is shown in Fig.~\reffig{fig:cmratio_fwb_ravg_uwerr_w_fit_comparison}.
In Fig.~\reffig{fig:cmratio_fwb_ravg_uwerr_w_fit_comparison} we can indeed identify a region bounded by $t/a\sim 4$ and
$t/a \sim 10$, where we find a dominating linear dependence on the source-sink time separation.

We list the results for the fit parameters for different choices of $f_{1/2}$ and fit intervals in Table~\ref{tab:fits}.
\begin{center}
\begin{table}
\caption{Parameters extracted using $f_1$, $f_2$ with no cubic term, and $f_2$
  using different fits ranges $t_{\rm min}/a$ to
$t_{\rm max}/a$. In the last column we give the $\chi^2$ per degree of freedom (dof).}
\begin{tabular}{ccccccc}
  & $t_\mathrm{min}/a$ & $t_\mathrm{max}/a$ & $A\cdot 10^{2}$ & $B\cdot 10^2$ & $C\cdot 10^5 / a\delta$ & $\chi^2/\mathrm{dof}$ \\
\hline
  $f_1$ &  $4$ & $ 9$ & $6.47 \,( 49 )$ & $ 2.62 \,(15 ) $ & $ 0.188 \,( 68 )$ & $ 2.4 / 3$\\
  $f_1$ &  $4$ & $10$ & $6.24 \,( 47 )$ & $ 2.69 \,(14 ) $ & $ 0.156 \,( 79 )$ & $ 4.3 / 4$\\
  $f_1$ &  $5$ & $ 9$ & $5.62 \,(103 )$ & $ 2.82 \,(26 ) $ & $ 0.140 \,(104 )$ & $ 1.8 / 2$\\
  $f_1$ &  $5$ & $10$ & $5.05 \,( 84 )$ & $ 2.98 \,(21 ) $ & $ 0.074 \,(122 )$ & $ 2.9 / 3$\\
\hline
  $f_2$ &  $4$ & $ 9$ & $5.62 \,(25 )$ & $ 2.89 \,(06 ) $ & & $ 6.0 / 4$\\
  $f_2$ &  $4$ & $10$ & $5.63 \,(25 )$ & $ 2.89 \,(06 ) $ & & $ 6.5 / 5$\\
  $f_2$ &  $5$ & $ 9$ & $4.75 \,(51 )$ & $ 3.05 \,(10 ) $ & & $ 2.4 / 3$\\
  $f_2$ &  $5$ & $10$ & $4.78 \,(52 )$ & $ 3.05 \,(11 ) $ & & $ 3.0 / 4$\\
\hline
  $f_2$ &  $4$ & $ 9$ & $6.51 \,( 53 )$ & $ 2.60 \,(16 )$ & $ 4.1 \,(22 )$ & $ 2.4 / 3$\\
  $f_2$ &  $4$ & $10$ & $6.27 \,( 52 )$ & $ 2.68 \,(16 )$ & $ 2.9 \,(21 )$ & $ 4.3 / 4$\\
  $f_2$ &  $5$ & $ 9$ & $5.64 \,(128 )$ & $ 2.82 \,(33 )$ & $ 2.4 \,(32 )$ & $ 1.8 / 2$\\
  $f_2$ &  $5$ & $10$ & $5.05 \,(117 )$ & $ 2.98 \,(30 )$ & $ 0.7 \,(28 )$ & $ 2.9 / 3$\\
\hline\hline
\label{tab:fits}
\end{tabular}
\end{table}
\end{center}

An important outcome is that the value for the slope  $B$ is stable 
when using different fits ranges and the two fitting ans\"atze. 
We also find a positive value for the energy splitting, as expected, although the statistical error is large not allowing a precise determination
of the splitting $\delta$ from these fits. This is a consequence of 
 the observation that already the linearized fit gives a satisfactory description of the data as indicated by the value of the
 $\chi^2/\mathrm{dof}$. Although allowing
for deviations from the linear dependence  tends to decrease  the value extracted for the slope, it also leads to an increase
of the statistical uncertainty.
In fact, the difference between  the values of the  slopes determined using $f_1$ and $f_2$ agree within two standard deviations.
The variation of the fit parameters with the lower boundary
of the fit interval
probes the sensitivity to the influence of excited states. As can be seen,
the changes in the slope are within the statistical uncertainty and thus
to the accuracy of our measurements we do not see excited state contamination.

\section{Extraction of the coupling}
The value of the slope $B$ is connected with the asymptotic behavior of the correlator at large Euclidean time.
In this limit we expect the slope to be associated with the following expression:
\ba
B_i &=& \sum\limits_{\sigma_3,\,\tau_3}\,\frac{\matelem(\Qvec,\,\qvec,\,\sigma_3,\,\tau_3)}{\sqrt{N_\Delta\,N_{\pi N}}}\,V\,\delta_{\Qvec \Qvec}
  \frac{\Xi_i(\sigma_3,\,\tau_3)}{\sqrt{\Sigma^\Delta_i\,\Sigma^N}}\,,
\ea
where
$\Sigma^{\Delta}_{i}$ and $\Sigma^N$ denote the spin sums arising from the corresponding 2-point functions of the
$\Delta$ and the nucleon in the denominator. We include the index $i$ denoting the component of the momentum vector $\qvec$,
which is non-zero. We also have
\ba
  \Xi_i(\sigma_3,\,\tau_3) &=& \Gamma^{(4)}_{\beta\alpha}\,u_\Delta^{i\,\alpha}(\Qvec=0,\,\sigma_3)\,u_N^{\beta}(\qvec,\,\tau_3)\,.\nonumber
\ea
We use the standard normalization for the fermionic and bosonic states given by
\ba
N_\Delta &=& V\,\frac{E_\Delta}{m_\Delta} \\
N_{\pi N} &=& N_\pi \times N_N = 2V\,E_\pi \times V\,\frac{E_N}{m_N}\,.
\ea
Note that in accordance with our approximation of the pion-nucleon bound state we normalize the latter as a product
of single particle states.
The factor $V\,\delta_{\Qvec\Qvec}$ with a lattice Kronecker-$\delta$ reflects the conservation of spatial momentum for
our lattice kinematics, where we match exactly $\pvec_\Delta = \pvec_{\pi N}$.
Finally, $\matelem$ is the transition matrix element, which to leading order we connect using  effective field
theory~\cite{Pascalutsa:2005vq} to the coupling constant by the relation
\ba
\matelem(\Qvec,\,\qvec,\,\sigma_3,\,\tau_3) &=& \frac{g^{\Delta}_{\pi N}}{2m_N}\,\ubar_\Delta^{\mu\,\alpha}(\Qvec,\sigma_3)\,q_\mu\,u_N^{\alpha}(\Qvec+\qvec,\tau_3)\,,
\ea
where we consider  the isospin channel $I_3 = +3/2$.
With our specific choice $\Qvec = 0$ and $\qvec \propto \hat{e}_i$ we find
\ba
B_i\,\frac{\sqrt{N_\Delta\,N_{\pi N}}}{V} &=& g^{\Delta}_{\pi N}\,\frac{q_i}{2m_N}\,\sqrt{\frac{1}{3}\,\frac{E_N + m_N}{m_N}}\,.
\ea
We solve this relation to extract the coupling $g_{\pi N}^\Delta$. Combining the results from all fits we find
\ba
g_{\pi N}^\Delta(\mathrm{lat}) &=& 27.0 \,(6)\,(15)\,.
\label{lattice value}
\ea
We estimate the systematic error from the variance of the results for $g_{\pi N}^\Delta$ from the individual fits.

\section{Discussion and Outlook}

The method outlined here is based on the mixing of hadronic states on the lattice provided their energies are close. This enables us to compute the overlap of these states even if the particle is above the decay threshold. 
This overlap can then be related to the coupling constant by connecting to the
effective field  theory to leading order.  
Although we have only obtained results 
 for one lattice spacing and volume, our previous studies of the properties of these hadrons have not shown large lattice artifacts and therefore we expect the same to hold true for this calculation~\cite{Alexandrou:2010uk, Alexandrou:2013opa}. The pion mass dependence of the $\pi N \Delta$ form factor at non-zero 
momentum transfer was studied using DW fermions with smallest pion mass about 300~MeV and with the same hybrid ensemble as this work~\cite{Alexandrou:2010uk}. Within this range of pion mass
no large variation was observed. However, one would need to perform the same calculation
closer to the physical pion mass in order to access the pion mass dependence
of the coupling. 
 Nevertheless we can compare with other determinations
bearing in mind that our value holds for a pion mass of about 360~MeV.

First, let us look at the result for the width in leading order continuum effective field theory (cf. \cite{Pascalutsa:2005vq}),
\ba
\Gamma &=& \frac{g_{\pi N \Delta}^2}{48\pi}\,\frac{1}{m_N^2}\,\frac{E_N + m_N}{E_N + E_{\pi}}\,q^3\,.
\ea
Together with the PDG value for the width $\Gamma = 118\,(3)\mev$ this leads to a coupling
\ba
g_{\pi N}^\Delta(\mathrm{lo~eft}) &=& 29.4\,(4)\,.
\ea
Secondly, in Ref.~\cite{Hemmert:1994ky}  an experimental value was derived
 based on a model-independent $K$-matrix analysis, which reads
\ba
  g_{\pi N}^\Delta(\mathrm{exp}) &=& 28.6\,(3)\,.
\ea
We find that our result is compatible with both these values, which is remarkable given that this is a determination at higher than physical pion mass.

In order to calculate the width we will consider the continuum expressions. We note here that 
 the width cannot be calculated by simply using lattice results in e.g. the 
leading order effective field theory formula since the lattice setup differs
from  the physical decay process. The problem is rooted
in the non-conservation of energy in the lattice setup. Note for instance, that for an experimental decay, the relative
momentum in the pion-nucleon system is fixed at $k_\mathrm{exp} \approx 227\mev$, while in our lattice setup with
$a^{-1} \approx 1.6\gev$ one unit of momentum corresponds to a much larger value $k_\mathrm{lat} \approx 360\mev$. 
 Assuming a flat scaling of the dimensionless $g_{\pi N}^\Delta(\mathrm{lat})$ given in Eq.~(\ref{lattice value}), neglecting volume dependence and using the continuum relation
we  obtain an estimate of the decay width of the $\Delta$:
\ba
  \Gamma_\Delta &=& 99\,(12)\mev\,.
\ea
An alternative method to evaluate the width is based on the L\"uscher approach 
that measures the energies as a function of the lattice
spatial length L. Although this method has been applied successfully to calculate
the decay width of mesons~\cite{Mohler:2012nh} application to baryon decays is still
limited by the accuracy attainable in baryon systems. The generalization of L\"uscher's formulation in the case of the $\Delta$ has been given in Ref.~\cite{Gockeler:2012yj} and some preliminary results have been obtained~\cite{Mohler:2012nh}. A matrix Hamiltonian method applicable to $\Delta \to \pi N$ has also been developed~\cite{Hall:2012wz}.
The method presented here is thus a valuable alternative that
yields a reliable result for the coupling constant, which can then be
related to the width using continuum relations.
This first application to the $\Delta$ width has demonstrated the applicability
of the method.  Future work will aim at quantifying systematic uncertainties  
by performing  analyses
for different lattice spacings, spatial volumes and pion masses as well as
studying 
the impact of a
pion-nucleon state beyond the non-interacting case.

The method presented here can immediately be applied to a number of other hadronic decays, such as
$\Sigma^{\frac{3}{2}+} \to \Lambda \pi$ or
$\Sigma^{\frac{1}{2}-} \to \bar{K} N$ as well as for the corresponding charmed baryons.
A calculation of the widths of these baryons is feasible using this method and work in this direction
is underway.
Moreover, given the statistical accuracy achieved for the coupling compared to the systematic error,
even with the moderate 
ensemble size used for this work, tackling excited baryonic state decays may become feasible within the same approach.

\noindent
{\bf Acknowledgments:}
{\small We would like to thank C. Michael for valuable discussions.
This research was in part supported by the Research Executive Agency of the 
European Union under Grant Agreement number PITN-GA-2009-238353 (ITN STRONGnet)
and in part by the DOE Office of Nuclear Physics under grant \#DE-FG02-94ER40818.
The GPU computing resources were provided by the Cy-Tera machine at the Cyprus Institute supported in
part by the Cyprus Research Promotion Foundation  under contract
  NEA Y$\Pi$O$\Delta$OMH/$\Sigma$TPATH/0308/31, the National Energy Research Scientific 
Computing Center supported by the Office
of Science of the DOE under Contract No. DE-AC02-05CH11231 and 
by the J\"ulich Supercomputing Center, awarded under the PRACE EU FP7 project 2011040546.
The multi-GPU domain wall inverter code~\cite{Strelchenko:2012aa} is based on the QUDA library~\cite{Clark:2009wm,Babich:2011np}
and its development has been supported by PRACE grants RI-211528 and FP7-261557.
}

\bibliography{letter_3}
\bibliographystyle{h-physrev}
\end{document}

%% file: adjustment.tex
\usepackage{amsmath}
\usepackage{slashed}
\usepackage{amscd}
\usepackage{amssymb}
\usepackage{latexsym}
\usepackage{verbatim}
\usepackage{mathbbol}
\usepackage{graphicx}
\usepackage{color}
\usepackage{hyperref}

\newcommand{\brackets}[1]{\langle #1 \rangle}

\newcommand{\epow}[1]{\mathrm{e}^{#1}}

\newcommand{\gammafive}{\gamma_{5}}

\newcommand{\mps}{m_{PS}}

\newcommand{\fermi}{\,\mathrm{fm}}
\newcommand{\gev}{\,\mathrm{GeV}}
\newcommand{\mev}{\,\mathrm{MeV}}

\newcommand{\ubar}{\bar{u}}
\newcommand{\dbar}{\bar{d}}

\newcommand{\balign}{\begin{align}}
\newcommand{\ealign}{\end{align}}
\newcommand{\beq}{\begin{equation}}
\newcommand{\eeq}{\end{equation}}
\newcommand{\balignat}[1]{\begin{alignat}{#1}}
\newcommand{\ealignat}{\end{alignat}}

\newcommand{\bfig}{\begin{figure}}
\newcommand{\efig}{\end{figure}}

\newcommand{\bc}{\begin{center}}
\newcommand{\ec}{\end{center}}

\newcommand{\btab}{\begin{table}}
\newcommand{\etab}{\end{table}}

\newcommand{\bcom}{}

\newcommand{\bitem}{\begin{itemize}}
\newcommand{\eitem}{\end{itemize}}

\newcommand{\benum}{\begin{enumerate}}
\newcommand{\eenum}{\end{enumerate}}

\newcommand{\pvec}{\vec{p}}

\newcommand{\qvec}{\vec{q}}
\newcommand{\Qvec}{\vec{Q}}
\newcommand{\xvec}{\vec{x}}
\newcommand{\yvec}{\vec{y}}

\newcommand{\Deltapp}{\ensuremath{\Delta^{++}}}

\newcommand{\matelem}{\mathcal{M}}

\newcommand{\reffig}[1]{[\ref{#1}]}